\documentclass[journal]{IEEEtran}
\usepackage{amsfonts,epsfig, cite ,array,graphicx,amsmath,amssymb,epstopdf}
\usepackage{longtable, makecell}
\usepackage{multirow}
\usepackage{amsmath}
\usepackage{booktabs}
\usepackage{capt-of}
\usepackage{balance}
\usepackage{subfigure}
\usepackage{setspace}
\usepackage{color, colortbl}
\usepackage{adjustbox}
\usepackage{hyperref}
\usepackage{mathtools}
\usepackage{svg}
\usepackage{graphicx}

\usepackage{algorithm}
\usepackage{algpseudocode} 

\begin{document}
\title{Working Memory Functional Connectivity Analysis for Dementia Classification using EEG}

\author{Shivani Ranjan$^1$, Anant Jain$^1$, Robin Badal$^2$, Amit Kumar$^3$, Harshal Shende$^1$, Deepak Joshi$^3$, Pramod Yadav$^2$, and Lalan Kumar$^{1}$*, 



\thanks{*Correspondence: lkumar@ee.iitd.ac.in}
\thanks{$^1$Department of Electrical Engineering, Indian Institute of Technology Delhi, India}
\thanks{$^2$Department of RS and BK, All India Institute of Ayurveda Delhi, India.}
\thanks{Full list of author information is available at the end of the article}}

\maketitle
\begin{abstract}
\newline \textit{Background:} Dementia, particularly Alzheimer’s Disease (AD), is a progressive neurodegenerative disorder marked by cognitive decline. Early detection, especially at the Mild Cognitive Impairment (MCI) stage, is essential for timely intervention. Working Memory (WM) impairment is a key early indicator of neurodegeneration, affecting higher cognitive processes. Electroencephalography (EEG), with its high temporal resolution, offers a cost-effective method to assess brain dynamics. This study investigates WM-related EEG functional connectivity (FC) to identify brain network alterations across dementia stages.
\newline \textit{Methods:} EEG signals were recorded from 24 participants (8 AD, 8 MCI, and 8 healthy controls) during WM tasks, including encoding, recall, and retrieval stages. Data preprocessing involved noise reduction and feature extraction using Spherical and Head Harmonic Decomposition (SHD, HHD). FC was quantified using Cross-Plot Transition Entropy (CPTE) and Phase Lag Index (PLI). Network metrics such as Degree and Eigenvector Centrality were analyzed using Support Vector Machine, Random Forest, and XGBoost classifiers.
\newline \textit{Results:} The CPTE-based connectivity metrics outperformed the traditional PLI approach in differentiating dementia stages, attaining a peak classification accuracy of 97.53\% during the retrieval phase with the Random Forest model. A connectivity threshold of 0.5 was optimal for network discrimination. SHD and HHD features also demonstrated strong discriminative potential. AD subjects exhibited higher synchronization patterns during WM tasks than healthy controls.
\newline \textit{Conclusions:} The integration of WM tasks with EEG-based FC analysis provides a robust framework for dementia classification. The proposed CPTE-based approach offers a robust, scalable, non-invasive, and effective diagnostic tool for early detection and monitoring of neurodegenerative diseases.

\end{abstract}


\section{Introduction}\label{sec:intro}

\subsection{Background \& Related Work}\label{subsec:literature}

Alzheimer's disease (AD) is a progressive neurodegenerative disorder and the leading cause of dementia. With its rising prevalence and the lack of effective treatments to halt \cite{ang2020vascular,cutsuridis2017computational,warren2021efficacy} or reverse its progression \cite{rasmussen2019alzheimer,hill2020longitudinal}, AD has become one of the most burdensome and costly diseases of this century \cite{jia2018cost}. The condition is characterized by impairments in working memory, episodic memory, and executive function, that significantly impact cognitive abilities and daily living. The symptoms typically manifest in individuals over 65 years old, however, early-onset AD can begin as early as 30 years of age. This presents unique challenges in diagnosis and management \cite{kirova2015working,ghayedi2024review,dubois2016preclinical}. Mild cognitive impairment (MCI) is considered an intermediate stage between normal aging and AD, offering a critical period for intervention \cite{petersen2011clinical}. It is characterized by subjective memory complaints and measurable cognitive decline, with daily functioning largely preserved \cite{egerhazi2007automated,petersen2005mild,winblad2004mild}. This stage presents an opportunity to implement strategies to slow disease progression \cite{kirova2015working}. Approximately 50\% of individuals with MCI progress to AD within five years \cite{tabatabaei2020cognitive}, underscoring the urgency of early and accurate detection to mitigate the long-term effects of neurodegeneration \cite{reinvang2012executive}.  

Neurocognitive assessments are widely employed to evaluate cognitive skills in dementia patients \cite{morris1993clinical,folstein1975mini,nasreddine2005montreal,bruno2019addenbrooke}. These tools rely on questionnaires designed to measure functional abilities and provide valuable diagnostic insights for clinicians. However, their effectiveness is often limited by several factors. They are sensitive to variations in education level and premorbid intelligence, which can bias the results. Additionally, these assessments require extensive rater certification and depend heavily on expert clinical judgment during both administration and scoring. The process is also time-intensive. It typically takes around 30 minutes to complete. These limitations underscore the need for alternative, efficient methods that can enhance diagnostic accuracy while reducing dependency on subjective evaluations and resource-intensive procedures \cite{mendez2022chapter, bruno2019addenbrooke}.

Working memory (WM) decline is a prominent early indicator of neurodegeneration in MCI and AD \cite{jiang2024exploration,ghayedi2024review}. WM tasks involve the retention and manipulation of information and play a critical role in cognitive functions such as decision-making, problem-solving, and learning. Neuroimaging-based studies, including functional magnetic resonance imaging (fMRI) and positron emission tomography (PET), have consistently shown reduced activation in the frontal and parietal regions during WM tasks in MCI patients \cite{wang2024altered,lou2015decreased,li2023functional,ghayedi2024review}. These findings suggest that disruptions in brain activity in these regions are closely linked to cognitive decline. Structural and functional changes in key memory-related areas, such as the hippocampus and the default mode network \cite{jones2016cascading}, have also been associated with disease progression \cite{ghayedi2024review}. These alterations are believed to reflect the breakdown of neural circuits essential for higher-order cognitive functions \cite{jones2016cascading}. Despite the insights provided by fMRI and PET, these brain imaging modalities face practical limitations in clinical use. Their high cost, computation time, lack of portability, and sensitivity to motion artifacts make them unsuitable for clinical diagnostics and large population studies. This has necessitated the exploration of alternative, cost-effective, and accessible methods for assessing brain function and tracking WM-related changes in neurodegenerative conditions.    

Electroencephalography (EEG) offers a practical, non-invasive, and cost-effective method for assessing brain function \cite{jain2023subject}. Its high temporal resolution enables real-time monitoring of neural activity, providing valuable insights into the brain dynamics \cite{jain2025esi}. EEG-based functional connectivity (FC) analysis has been effective in detecting disruptions in brain network integration caused by neurodegeneration \cite{ranasinghe2024functional,ranjan2024exploring}. In previous studies, resting-state EEG has demonstrated the changes in connectivity across frequency bands (delta, alpha, beta, and gamma bands) and its correlation with cognitive decline in conditions such as AD and MCI \cite{meghdadi2021resting,ranjan2024exploring,ranjan2024dementia}. EEG studies on WM have also been conducted in individuals with cognitive disorders. These studies reveal that the changes in EEG power or specific frequency bands reflect brain function under high WM loads \cite{eskikurt2024evaluation,yu2024eeg}. However, the brain network dynamics during different WM stages—encoding, retro cue, recall, and retrieval remain underexplored.

\subsection{Objectives and Contributions}\label{subsec:contribution}

In this study, a comprehensive framework based on surface EEG-derived brain network organization metrics is proposed to classify AD, MCI, and healthy controls (HC) across different WM stages. The EEG dataset comprises recordings from HC, preclinical AD (MCI), and AD participants performing WM tasks. A robust and noise-tolerant synchronization measure, Cross-Plot Transition Entropy (CPTE), is utilized to construct the Network Organization Matrix (NOM) and characterize brain network dynamics. Additionally, low-dimensional spatiotemporal EEG features derived from spherical and head harmonics are explored for their potential in dementia classification. Multiple machine learning classifiers are employed to evaluate classification performance across the encoding, retro-cue, recall, and retrieval stages of WM.

The key objectives of this study are as follows:
\begin{itemize}

    \item To collect and analyze EEG data from HC, MCI, and AD participants for investigating stage-specific neural dynamics of neurodegeneration.
    \item To employ CPTE-based NOM for analyzing brain network organization during WM tasks.
    \item To extract spherical and head harmonics–based EEG features for dementia classification.
    \item To determine the most informative WM stage (encoding, retro-cue, recall, retrieval) for accurate classification.
    \item To systematically evaluate the discriminative power of EEG features, NOMs, and network parameters using machine learning classifiers.
\end{itemize}


The organization of the article is as follows. Section \ref{sec:expsetup} includes the description of the experimental setup and recording paradigm. Section \ref{sec:material} details the preprocessing pipeline of EEG data (Section \ref{subsec:prepro}), harmonics decomposition (Section \ref{subsec:harmonics}), computation of NOM (Section \ref{subsec:NOM_compute}), and extraction of network parameters (Section \ref{subsec:network_para}). The findings of the study are presented in Section \ref{sec:results}, while Section \ref{sec:discuss} discusses the interpretation of the results. Section \ref{sec:conclusion} provides the conclusion of the research study. 
\begin{figure*}[ht]
  \centering
  \includegraphics[width = 0.9\linewidth]{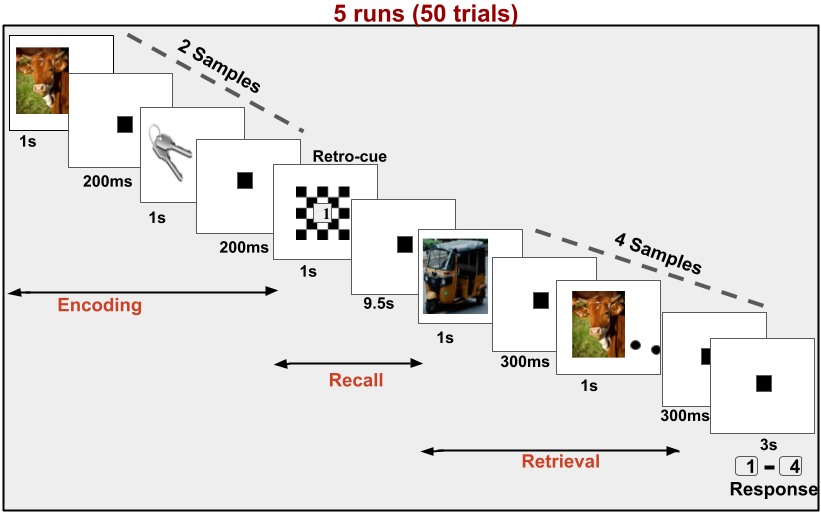}
  \caption{Recording paradigm of the retro-cue based match-to-sample task over an extended delay period.}
  \label{fig:EP_BD}
\end{figure*}

\section{Experimental Setup and Data Acquisition}\label{sec:expsetup}
This Section includes the description of participants, recording equipment, experimental protocol, and task paradigm to analyze working memory EEG data for dementia classification.

\subsection{Participants and Equipments}\label{subsec:par&equip}

The study incorporated twenty-four right-handed participants (AD = 8; MCI = 8; HC = 8, age range: 60-80 years) of Indian origin with normal or corrected-to-normal vision. The data collection protocol was approved by the Institute Ethics Committee, All India Institute of Ayurveda, New Delhi. The diagnosis of AD, MCI, and HC groups was established using Mini-Mental State Examination (MMSE) \cite {lacy2015standardized} and Montreal Cognitive Assessment (MoCA) screening tools \cite{freitas2013montreal}. The participants with consistent categorization across both the scales (MMSE: AD $<$ 18, MCI 18-25, HC $>$ 25; MoCA: AD $<$ 21, MCI 21-26, HC $>$ 26) were included in the analysis. HC group participants reported no history of neurological or psychiatric disorders. All participants provided their informed consent prior to the study. EEG data was recorded using a 64-channel Ag/AgCl gel-active electrode EEG system (actiCHamp, Brain Products GmbH, Germany) with Fz as the reference electrode. EEG electrodes were placed according to the 10:10 EEG electrode placement system, and impedance resistance was maintained below 10 k$\Omega$ using conductive gel, ensuring a high signal-to-noise ratio. The EEG signals were recorded at a sampling rate of 1000 Hz without any internal filters.

\subsection{Experimental Protocol}\label{subsec:exp_paradigm}

The experiment protocol was designed to capture three distinct WM stages: encoding, recall (delay or maintenance), and retrieval. The recording paradigm of the experiment is illustrated in Figure \ref{fig:EP_BD}. Each trial initiated with the encoding phase in which the participant was instructed to memorize a set of 2 images (termed as samples). Each sample stimulus was shown for 1 s, followed by a fixation period of 200 ms. Later, a retro-cue (‘1’ or ’2’) was presented on a black and white checkerboard background for 1000 ms, instructing participants to remember the corresponding sample. This was followed by the presentation of a blank screen (with only the fixation point) for a recall period of 9.5 s, resulting in 10.5 s of overall delay. In retrieval stage, four probe images were displayed sequentially, each for 1000 ms with a 300 ms inter-stimulus time interval. The participants were instructed to identify the retro-cued sample image among probe images and respond using a keyboard button (keys 1-4)
with their right hand. Numbers 1-4 indicate the probe image sequence. The participant presses the key matching the retro-cued image. After the offset of the fourth probe image screen, the response time was limited to 3000 ms. This variant of the match-to-sample task required participants to identify the memorized sample out of a larger set of stimuli (chance level: 25$\%$) to minimize ceiling effects \cite{yan2021decoding}. Each participant underwent 20 practice trials with non-experimental stimuli to familiarize them with the task paradigm. The stimuli utilized in the study consisted of simplified and culturally familiar images to ensure comprehensibility across the participants. Each image was unique for within the trial run unless reused as a probe.

All experiments were conducted in a quiet and dimly lit room with participants sat comfortably on a chair. A user interface was developed using Python's Tkinter module to present the stimuli on a PC laptop and record the responses using a keyboard. Every participant in each group completed an EEG recording of 5 runs, with 10 randomized trials per run. The total duration of a run was (21.1s x 10) 3.5 minutes, followed with a variable inter-run interval of 20–50 s to prevent fatigue. Thus, for each participant, the number of sessions for encoding, retro-cue, recall, and retrieval were 100 (of 1 s), 50 (of 1 s), 50 (of 9.5 s), and 200 (of 1 s), respectively. Hence, total sessions across each study group for encoding, retro-cue, recall, and retrieval were 800, 400, 400, and 1600, respectively. The response was missed multiple times by each group, especially the AD study group.

\section{Materials and Methods}\label{sec:material}
This Section details the proposed methodology employed for dementia classification using EEG data recorded during working memory tasks. It consists of EEG pre-processing, spherical and head harmonic decomposition, computation of network organization matrices (NOMs), and network parameters extraction for input to classifiers in Sections \ref{subsec:prepro} - \ref{subsec:network_para}.

\subsection{EEG Preprocessing}\label{subsec:prepro}
EEG data preprocessing was performed for noise and artifact removal to ensure accurate connectivity and network analysis. The preprocessing steps were conducted using EEGLAB \cite{delorme2004eeglab} plugin of the MATLAB 2022b software. Continuous EEG data was bandpass filtered between 0.5 Hz and 40 Hz using a fourth-order Butterworth filter to remove low-frequency slow drifts. Independent component analysis (ICA) was applied to identify and remove the components associated with ocular and muscle activities if their variance exceeded a threshold of 0.6. The preprocessed EEG data were re-referenced to the average reference. Further data were segmented into 1 s epochs corresponding to each WM stage (encoding, retro-cue, recall, and retrieval).

\subsection{Harmonics Decomposition}\label{subsec:harmonics}

The spherical and head harmonics decomposition (SHD/HHD) for EEG was proposed in \cite{giri2020brain,giri2022anatomical} for efficient data representation and dimensionality reduction. This is useful when high-density EEG is utilized. In this study, the coefficients obtained from SHD and HHD are additionally utilized in CPTE computation for dementia classification.


\subsubsection{Spherical Harmonics Decomposition (SHD)}
The human head is often approximated as a sphere in literature. This assumption allows the use of spherical harmonics as basis functions for spatial representation and sampling of scalp-recorded signals.

\begin{figure}[t]
    \centering
     \adjustbox{trim= 2cm 3.5cm 2.5cm 0}{%
    \includegraphics[width = 1 \linewidth]{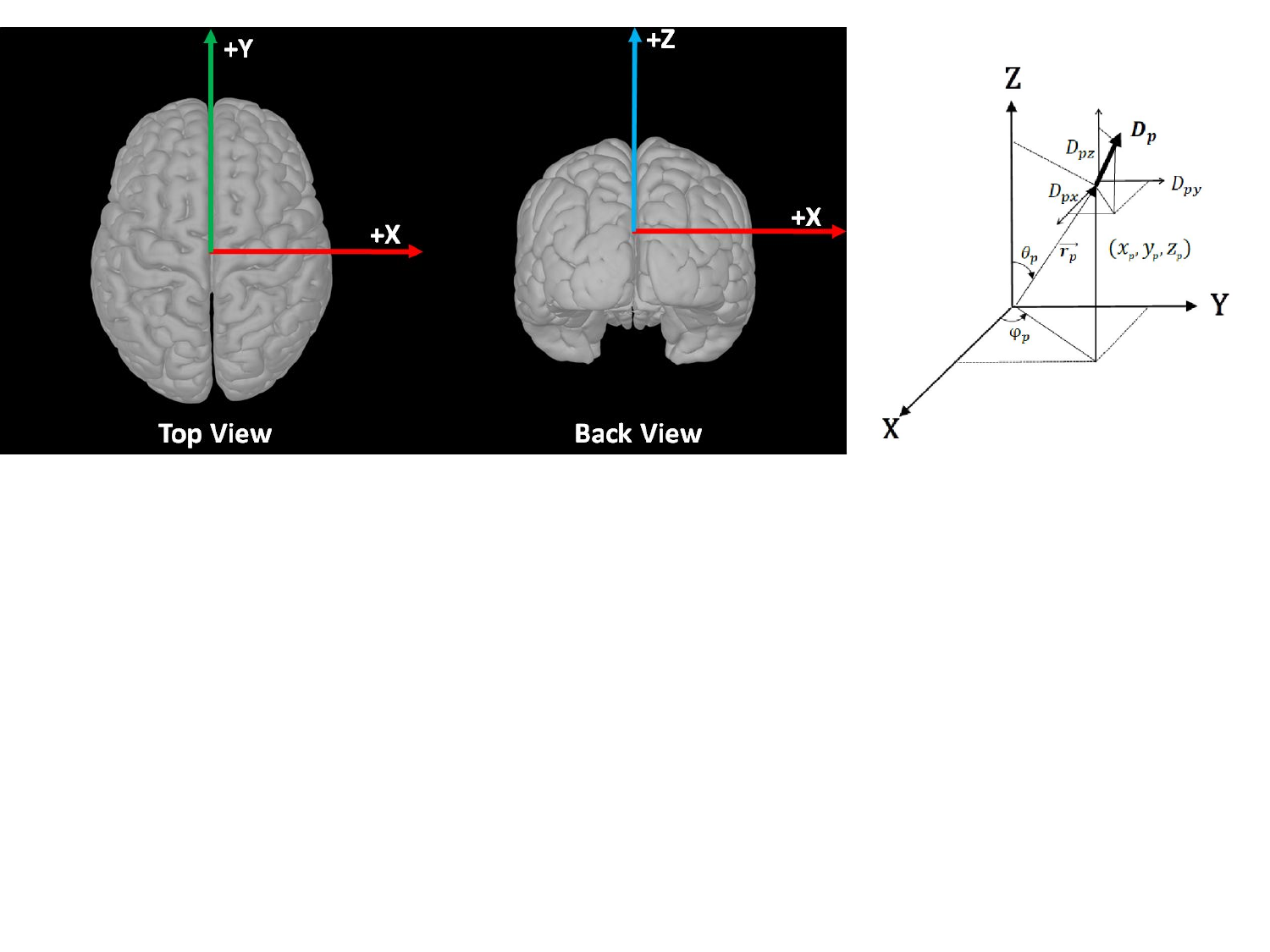}}
    \caption{Human head geometry with perimeter = 40 cm and radius = 10 cm.}
    \label{fig:head_geo}
\end{figure}

At time instant \( t \), the EEG potential at the location \( \mathbf{\Omega} = (R, \theta, \phi) \), generated by a dipole source located at \( \mathbf{\Omega_p} = (r_p, \theta_p, \phi_p) \), is represented as \( V(\mathbf{\Omega}, \mathbf{\Omega_p}, t) \). Here, \( R \), \( \theta \), and \( \phi \) correspond to the head radius, elevation angle, and azimuth angle, respectively, as per the coordinate measures shown in Figure \ref{fig:head_geo}. The SHD of the EEG potential can be expressed as \cite{giri2020brain}:

\[
V^{\text{SH}}_{nm} (\mathbf{\Omega_p}, t) = \int_{\mathbf{\Omega}} V(\mathbf{\Omega}, \mathbf{\Omega_p}, t) Y^m_n(\mathbf{\Omega}) \, d\mathbf{\Omega} \tag{3}
\]

In the aforementioned equation, \( Y^m_n(\mathbf{\Omega}) \) denotes the set of spherical harmonics of order \( n \) and degree \( m \). Assuming a finite decomposition order \( N \), the order of spherical harmonics takes values \( n \in [0, N] \), while the degree takes values \( m \in [-n, n] \). This results in a total of \( (N + 1)^2 \) distinct harmonics. The maximum limit on the order is governed by the equation \( N \leq (\sqrt{I} - 1) \), where \( I \) is the number of EEG sensors. In this study, 63 EEG electrodes are utilized for data recording, i.e., I = 63, and hence the upper limit of N is 6. Therefore, a total of 49 distinct spherical harmonics were obtained. 





In the discrete domain, the SHD in (3) can be rewritten as follows:
\[
V^{\text{SH}}_{nm} (\mathbf{\Omega_p}, t) = \sum_{i=1}^{I} \gamma_i V(\mathbf{\Omega_i}, \mathbf{\Omega_p}, t) Y^m_n(\mathbf{\Omega_i}) \tag{4}
\]

where, \( \mathbf{\gamma_i}\) denotes the sampling weight. The above equation can equivalently be written in matrix form for all \(n\) and \(m\) as follows:

\[
[V^{\text{SH}}_{nm}]_{(N+1)^2 \times T} = [Y]^T _{(N+1)^2\times I} \mathbf{\Gamma}_{I \times I} [V]_{I \times T} \tag{5}
\]

Here, \( \mathbf{\Gamma} \) is the collection of all the sampling weights, taken as the identity matrix in this study. The SHD of the processed EEG signal (\([V^{\text{SH}}_{nm}]\)) of each participant was extracted with the dimension of \((N + 1)^2 \times T\) as feature matrix. Here, \(T\) denotes the number of samples.
\begin{figure*}[t]
    \centering
    \includegraphics[width = 0.9\linewidth]{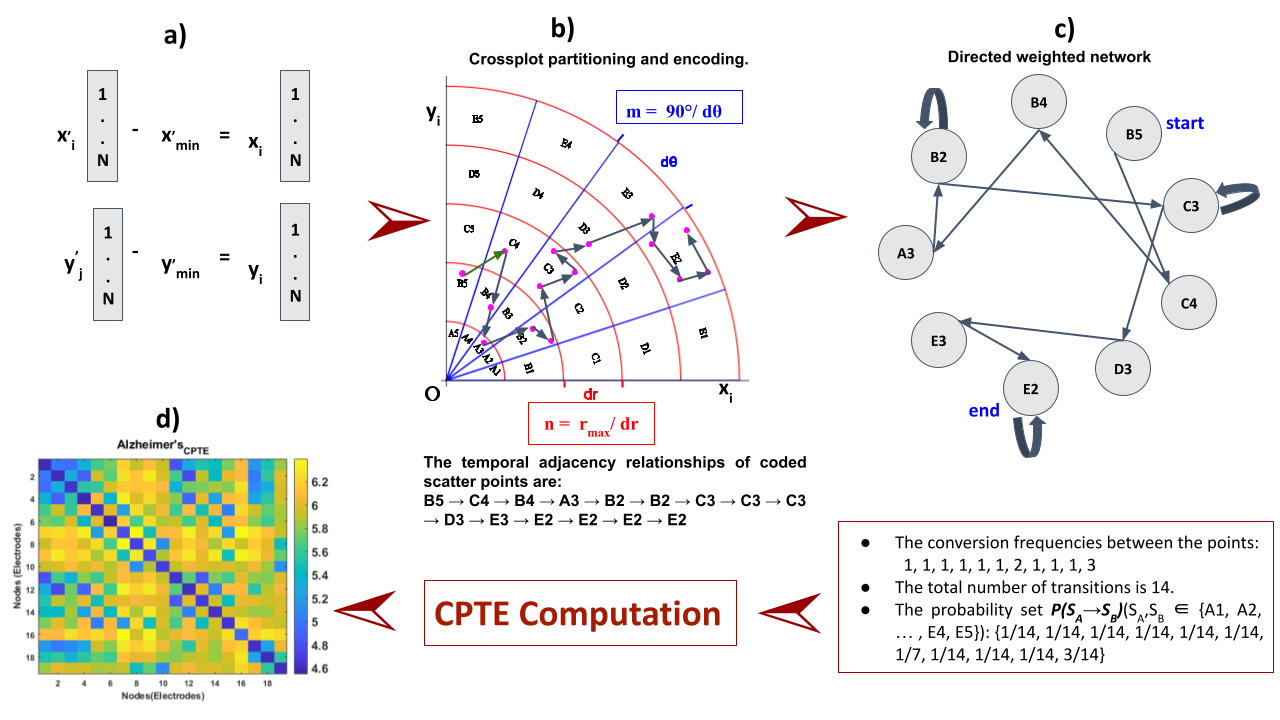}
    \caption{The diagram illustrates the computation of CPTE, where x'$_i$ and y'$_i$ represent two time series, in this case, an epoch of a possible electrode pair. In this analysis, the radial ruler ($dr$) and angular ruler ($d\theta$) are set to 10 and $10^\circ$, respectively. Here x'$_i$ and y'$_i$ correspond to $v_i$ and $v_j$}
    \label{CPTE_BD1}
\end{figure*}

\subsubsection{Head Harmonics Decomposition (HHD)} While the spherical head shape approximation provides a simplified framework, the actual geometry of the human head resembles a shape between a hemisphere and a sphere, as shown in Figure \ref{fig:head_geo}. Head harmonic basis functions are formulated for an accurate representation of data sampled over the human head \cite{giri2022anatomical,giri2020brain}. These head-specific basis functions better approximate the spatial distribution of EEG data over the scalp. The HHD coefficients are computed as:





\[
[V^{H^2}_{nm}] = [H]^T \mathbf{\Gamma} [V] \tag{6}
\]

Here, \( H^m_n(\mathbf{\Omega}) \) denotes the set of head harmonics of order \( n \) and degree \( m \). Similar to SHD, the dimension of the HHD  matrix is also \((N + 1)^2 \times T\). Since \((N + 1)^2 \leq I\) \cite{giri2020brain,giri2022anatomical}, the SHD and HHD reduce the dimension of EEG data while retaining the spatial information, thus improving computational efficiency.

\subsection{Network Organization Matrix (NOM) Computation}\label{subsec:NOM_compute}

This section has focused on the methodology for computing the network organization matrices (NOMs) to evaluate the effect of brain network organization in dementia classification. NOMs have been utilized to analyze the efficiency of the underlying cortical-electrical network structure. The Phase Lag Index (PLI) and Cross-Plot Transition Entropy (CPTE) have been employed to compute the NOM for each WM stage across AD, MCI, and HC groups. NOMs have been computed for extracted epochs \(e\) from \([V]\), \([V^{SH}_{nm}]\), \([V^{H^2}_{nm}]\) for AD, MCI, and HC groups and four WM stages.

 \subsubsection{Phase Lag Index (PLI)}

PLI measures phase synchronization between two EEG signals by evaluating the asymmetry of instantaneous phase differences between them. This reduces the impact of the volume conduction problem \cite{stam2007phase}. PLI values range from 0 to 1, where 0 indicates no coupling, and 1 denotes perfect phase locking. The PLI is expressed as:
\[
\text{PLI} = \left| \langle \text{sign}(\sin(\Delta \phi(t_k))) \rangle \right| \tag{7}
\]
where \( \Delta \phi(t_k) \) is the phase difference at time instant \( t_k\), and the sign function calculates the signum. \( \langle \cdot \rangle \) denotes the mean, and \( || \) indicates the absolute value. Instantaneous phases were extracted using the Hilbert transform with a 50 $\%$ sliding window overlap. 

\subsubsection{Cross-Plot Transition Entropy (CPTE)}

CPTE quantifies the regularity of ordinal pattern transitions in EEG signals using Shannon entropy to measure synchronization between them \cite{ranjan2024exploring,chen2023capturing}. CPTE computation is shown in Figure \ref{CPTE_BD1}. It is to be noted that the CPTE value is inversely proportional to the coupling strength between EEG signals. This method is robust for non-stationary and short-time series with a lower sensitivity to additive noise than traditional approaches. The CPTE is defined as:
\begin{equation}
    CPTE = - \sum P(\text{S}_\text{A} \rightarrow \text{S}_\text{B}) \log_2 P(\text{S}_\text{A} \rightarrow \text{S}_\text{B})  \tag{8}
\end{equation}
In the aforementioned equation, \( P(\text{S}_\text{A} \rightarrow \text{S}_\text{B}) \) is the discrete probability that represents the likelihood of a transition from state \( S_A \) to state \( S_B \), where \( S_A, S_B \in \{A1, A2, \dots, E4, E5\} \), based on the conversion frequencies in the directed weighted network. While calculating CPTE, the radial ruler $dr$ was set to 2.

The sliding time window method was applied to analyze each epoch, with a window length of 500 ms (500 samples) and a step size of 250 ms (250 samples, i.e., 50\% overlap). For each sliding window, NOM(v\(_i\), v\(_j\)) between electrode pairs, v\(_i\) and v\(_j\), was computed using synchronization metrics. Here, NOM(v\(_i\), v\(_j\)) represents the \((v_i, v_j)_{\text{th}}\) element of the synchronization matrix. For \([V]\), \([V^{SH}_{nm}]\), and \([V^{H^2}_{nm}]\), \(\frac{p(p-1)}{2}\) possible pairs were considered, where \( p \) is 63 for \([V]\), and \((N + 1)^2\) for \([V^{SH}_{nm}]\) and \([V^{H^2}_{nm}]\). Since NOM(v\(_i\), v\(_j\)) = NOM (v\(_j\), v\(_i\)), the NOM matrix is symmetrical, and the corresponding graph model is, therefore, undirected. Once the analysis of the entire recording of a participant was completed, each NOM symmetric matrix was normalized by Min-Max normalization to rescale the matrix elements in the range 0–1. The NOM matrices were computed for all WM stages and study groups. Further, the normalized NOM matrices were converted into binary adjacency matrices using a thresholding approach. This step involved varying the threshold values (in the range from 0.1 to 0.9, with a step size of 0.1). For each threshold, matrix elements greater than or equal to the threshold value were set to 1, indicating a connection, while elements below the threshold value were set to 0, effectively eliminating weaker connections. This binarization process emphasized the strongest functional connections in the brain network to facilitate the extraction of relevant network parameters. The resulting binary matrices were then used to compute network parameters such as degree, clustering coefficient, eigenvector centrality, betweenness centrality, and coreness centrality, which were subsequently employed as input features for machine learning classifiers.

\begin{table*}[t]
\centering
\caption{Mean accuracies for 3-class dementia classification by using different CPTE-based EEG connectivity features. The effect on accuracy by using different threshold values in conjunction with different classifiers is also depicted.}
\scalebox{0.8}{
\begin{tabular}{|c|c|ccccc|ccccc|ccccc|}
\hline
\multirow{2}{*}{\textbf{Events}}    & \multirow{2}{*}{\textbf{Threshold}} & \multicolumn{5}{c|}{\textbf{SVM}}                                                                                                                          & \multicolumn{5}{c|}{\textbf{RF}}                                                                                                                           & \multicolumn{5}{c|}{\textbf{XGBoost}}                                                                                                                      \\ \cline{3-17} 
                                    &                                     & \multicolumn{1}{c|}{\textbf{f1}} & \multicolumn{1}{c|}{\textbf{f2}}    & \multicolumn{1}{c|}{\textbf{f3}} & \multicolumn{1}{c|}{\textbf{f4}} & \textbf{f5} & \multicolumn{1}{c|}{\textbf{f1}} & \multicolumn{1}{c|}{\textbf{f2}} & \multicolumn{1}{c|}{\textbf{f3}} & \multicolumn{1}{c|}{\textbf{f4}}    & \textbf{f5} & \multicolumn{1}{c|}{\textbf{f1}} & \multicolumn{1}{c|}{\textbf{f2}} & \multicolumn{1}{c|}{\textbf{f3}} & \multicolumn{1}{c|}{\textbf{f4}}    & \textbf{f5} \\ \hline \hline
\multirow{9}{*}{\textbf{ENCODING}}  & 0.1                                 & \multicolumn{1}{c|}{37.96}       & \multicolumn{1}{c|}{39.45}          & \multicolumn{1}{c|}{37.29}       & \multicolumn{1}{c|}{39.62}       & 36.81       & \multicolumn{1}{c|}{40.15}       & \multicolumn{1}{c|}{40.29}       & \multicolumn{1}{c|}{40.11}       & \multicolumn{1}{c|}{40.42}          & 37.02       & \multicolumn{1}{c|}{39.63}       & \multicolumn{1}{c|}{40.22}       & \multicolumn{1}{c|}{39.46}       & \multicolumn{1}{c|}{40.32}          & 37.01       \\ \cline{2-17} 
                                    & 0.2                                 & \multicolumn{1}{c|}{38.86}       & \multicolumn{1}{c|}{46.16}          & \multicolumn{1}{c|}{39.13}       & \multicolumn{1}{c|}{46.16}       & 38.05       & \multicolumn{1}{c|}{50.71}       & \multicolumn{1}{c|}{51.39}       & \multicolumn{1}{c|}{50.45}       & \multicolumn{1}{c|}{51.82}          & 39.31       & \multicolumn{1}{c|}{48.64}       & \multicolumn{1}{c|}{51.21}       & \multicolumn{1}{c|}{48.55}       & \multicolumn{1}{c|}{51.14}          & 39.32       \\ \cline{2-17} 
                                    & 0.3                                 & \multicolumn{1}{c|}{41.56}       & \multicolumn{1}{c|}{55.94}          & \multicolumn{1}{c|}{40.59}       & \multicolumn{1}{c|}{55.65}       & 40.03       & \multicolumn{1}{c|}{66.67}       & \multicolumn{1}{c|}{69.15}       & \multicolumn{1}{c|}{66.40}       & \multicolumn{1}{c|}{69.87}          & 44.28       & \multicolumn{1}{c|}{60.44}       & \multicolumn{1}{c|}{67.81}       & \multicolumn{1}{c|}{61.24}       & \multicolumn{1}{c|}{67.82}          & 42.61       \\ \cline{2-17} 
                                    & 0.4                                 & \multicolumn{1}{c|}{43.86}       & \multicolumn{1}{c|}{66.18}          & \multicolumn{1}{c|}{43.18}       & \multicolumn{1}{c|}{66.91}       & 38.06       & \multicolumn{1}{c|}{80.14}       & \multicolumn{1}{c|}{85.33}       & \multicolumn{1}{c|}{80.27}       & \multicolumn{1}{c|}{86.80}          & 52.34       & \multicolumn{1}{c|}{69.41}       & \multicolumn{1}{c|}{82.37}       & \multicolumn{1}{c|}{70.10}       & \multicolumn{1}{c|}{82.07}          & 49.21       \\ \cline{2-17} 
                                    & 0.5                                 & \multicolumn{1}{c|}{51.15}       & \multicolumn{1}{c|}{67.75}          & \multicolumn{1}{c|}{47.44}       & \multicolumn{1}{c|}{76.00}       & 45.44       & \multicolumn{1}{c|}{85.26}       & \multicolumn{1}{c|}{94.24}       & \multicolumn{1}{c|}{87.14}       & \multicolumn{1}{c|}{\textbf{96.10}} & 64.44       & \multicolumn{1}{c|}{73.41}       & \multicolumn{1}{c|}{91.28}       & \multicolumn{1}{c|}{76.85}       & \multicolumn{1}{c|}{\textbf{92.57}} & 58.04       \\ \cline{2-17} 
                                    & 0.6                                 & \multicolumn{1}{c|}{60.64}       & \multicolumn{1}{c|}{71.71}          & \multicolumn{1}{c|}{57.84}       & \multicolumn{1}{c|}{78.25}       & 59.52       & \multicolumn{1}{c|}{85.66}       & \multicolumn{1}{c|}{94.34}       & \multicolumn{1}{c|}{89.00}       & \multicolumn{1}{c|}{95.57}          & 74.00       & \multicolumn{1}{c|}{76.55}       & \multicolumn{1}{c|}{91.66}       & \multicolumn{1}{c|}{82.28}       & \multicolumn{1}{c|}{92.01}          & 65.60       \\ \cline{2-17} 
                                    & 0.7                                 & \multicolumn{1}{c|}{62.72}       & \multicolumn{1}{c|}{79.55}          & \multicolumn{1}{c|}{70.90}       & \multicolumn{1}{c|}{80.55}       & 67.68       & \multicolumn{1}{c|}{83.88}       & \multicolumn{1}{c|}{93.36}       & \multicolumn{1}{c|}{90.26}       & \multicolumn{1}{c|}{93.94}          & 80.21       & \multicolumn{1}{c|}{78.78}       & \multicolumn{1}{c|}{90.54}       & \multicolumn{1}{c|}{86.46}       & \multicolumn{1}{c|}{90.71}          & 74.24       \\ \cline{2-17} 
                                    & 0.8                                 & \multicolumn{1}{c|}{61.72}       & \multicolumn{1}{c|}{\textbf{82.50}} & \multicolumn{1}{c|}{74.50}       & \multicolumn{1}{c|}{80.35}       & 68.57       & \multicolumn{1}{c|}{85.60}       & \multicolumn{1}{c|}{92.48}       & \multicolumn{1}{c|}{91.31}       & \multicolumn{1}{c|}{93.00}          & 84.47       & \multicolumn{1}{c|}{81.53}       & \multicolumn{1}{c|}{89.20}       & \multicolumn{1}{c|}{87.13}       & \multicolumn{1}{c|}{88.53}          & 79.63       \\ \cline{2-17} 
                                    & 0.9                                 & \multicolumn{1}{c|}{62.52}       & \multicolumn{1}{c|}{82.45}          & \multicolumn{1}{c|}{53.95}       & \multicolumn{1}{c|}{76.51}       & 67.17       & \multicolumn{1}{c|}{77.35}       & \multicolumn{1}{c|}{90.57}       & \multicolumn{1}{c|}{87.28}       & \multicolumn{1}{c|}{91.11}          & 85.51       & \multicolumn{1}{c|}{73.81}       & \multicolumn{1}{c|}{85.51}       & \multicolumn{1}{c|}{78.83}       & \multicolumn{1}{c|}{85.51}          & 81.17       \\ \hline \hline
\multirow{9}{*}{\textbf{RETRONUM}}  & 0.1                                 & \multicolumn{1}{c|}{37.56}       & \multicolumn{1}{c|}{40.37}          & \multicolumn{1}{c|}{37.42}       & \multicolumn{1}{c|}{40.17}       & 36.60       & \multicolumn{1}{c|}{40.71}       & \multicolumn{1}{c|}{40.94}       & \multicolumn{1}{c|}{40.71}       & \multicolumn{1}{c|}{41.28}          & 36.57       & \multicolumn{1}{c|}{40.40}       & \multicolumn{1}{c|}{40.88}       & \multicolumn{1}{c|}{40.31}       & \multicolumn{1}{c|}{40.94}          & 36.43       \\ \cline{2-17} 
                                    & 0.2                                 & \multicolumn{1}{c|}{40.00}       & \multicolumn{1}{c|}{48.11}          & \multicolumn{1}{c|}{40.23}       & \multicolumn{1}{c|}{47.83}       & 38.67       & \multicolumn{1}{c|}{50.89}       & \multicolumn{1}{c|}{51.92}       & \multicolumn{1}{c|}{50.87}       & \multicolumn{1}{c|}{52.40}          & 38.89       & \multicolumn{1}{c|}{49.02}       & \multicolumn{1}{c|}{51.66}       & \multicolumn{1}{c|}{49.16}       & \multicolumn{1}{c|}{52.14}          & 39.63       \\ \cline{2-17} 
                                    & 0.3                                 & \multicolumn{1}{c|}{42.98}       & \multicolumn{1}{c|}{54.89}          & \multicolumn{1}{c|}{42.64}       & \multicolumn{1}{c|}{55.89}       & 38.50       & \multicolumn{1}{c|}{66.58}       & \multicolumn{1}{c|}{69.25}       & \multicolumn{1}{c|}{66.24}       & \multicolumn{1}{c|}{70.27}          & 43.32       & \multicolumn{1}{c|}{60.74}       & \multicolumn{1}{c|}{69.42}       & \multicolumn{1}{c|}{61.73}       & \multicolumn{1}{c|}{68.00}          & 43.49       \\ \cline{2-17} 
                                    & 0.4                                 & \multicolumn{1}{c|}{43.75}       & \multicolumn{1}{c|}{66.38}          & \multicolumn{1}{c|}{42.44}       & \multicolumn{1}{c|}{66.61}       & 39.38       & \multicolumn{1}{c|}{77.90}       & \multicolumn{1}{c|}{85.30}       & \multicolumn{1}{c|}{78.41}       & \multicolumn{1}{c|}{86.81}          & 50.92       & \multicolumn{1}{c|}{69.33}       & \multicolumn{1}{c|}{83.35}       & \multicolumn{1}{c|}{70.41}       & \multicolumn{1}{c|}{83.94}          & 48.85       \\ \cline{2-17} 
                                    & 0.5                                 & \multicolumn{1}{c|}{49.62}       & \multicolumn{1}{c|}{62.07}          & \multicolumn{1}{c|}{45.33}       & \multicolumn{1}{c|}{72.03}       & 46.47       & \multicolumn{1}{c|}{82.07}       & \multicolumn{1}{c|}{93.28}       & \multicolumn{1}{c|}{84.23}       & \multicolumn{1}{c|}{\textbf{95.43}} & 62.78       & \multicolumn{1}{c|}{74.16}       & \multicolumn{1}{c|}{90.52}       & \multicolumn{1}{c|}{77.28}       & \multicolumn{1}{c|}{\textbf{93.25}} & 58.81       \\ \cline{2-17} 
                                    & 0.6                                 & \multicolumn{1}{c|}{60.17}       & \multicolumn{1}{c|}{70.07}          & \multicolumn{1}{c|}{56.77}       & \multicolumn{1}{c|}{78.58}       & 58.92       & \multicolumn{1}{c|}{83.35}       & \multicolumn{1}{c|}{94.01}       & \multicolumn{1}{c|}{87.60}       & \multicolumn{1}{c|}{94.95}          & 71.94       & \multicolumn{1}{c|}{77.16}       & \multicolumn{1}{c|}{91.57}       & \multicolumn{1}{c|}{82.72}       & \multicolumn{1}{c|}{92.82}          & 64.85       \\ \cline{2-17} 
                                    & 0.7                                 & \multicolumn{1}{c|}{62.30}       & \multicolumn{1}{c|}{80.57}          & \multicolumn{1}{c|}{70.07}       & \multicolumn{1}{c|}{82.35}       & 66.78       & \multicolumn{1}{c|}{82.33}       & \multicolumn{1}{c|}{93.76}       & \multicolumn{1}{c|}{90.33}       & \multicolumn{1}{c|}{93.70}          & 78.75       & \multicolumn{1}{c|}{79.72}       & \multicolumn{1}{c|}{90.24}       & \multicolumn{1}{c|}{86.44}       & \multicolumn{1}{c|}{90.44}          & 76.00       \\ \cline{2-17} 
                                    & 0.8                                 & \multicolumn{1}{c|}{61.64}       & \multicolumn{1}{c|}{\textbf{84.68}} & \multicolumn{1}{c|}{74.53}       & \multicolumn{1}{c|}{80.31}       & 69.42       & \multicolumn{1}{c|}{84.54}       & \multicolumn{1}{c|}{93.13}       & \multicolumn{1}{c|}{91.55}       & \multicolumn{1}{c|}{92.25}          & 83.94       & \multicolumn{1}{c|}{82.35}       & \multicolumn{1}{c|}{89.84}       & \multicolumn{1}{c|}{87.52}       & \multicolumn{1}{c|}{89.08}          & 80.85       \\ \cline{2-17} 
                                    & 0.9                                 & \multicolumn{1}{c|}{62.69}       & \multicolumn{1}{c|}{82.86}          & \multicolumn{1}{c|}{55.29}       & \multicolumn{1}{c|}{76.51}       & 68.51       & \multicolumn{1}{c|}{76.54}       & \multicolumn{1}{c|}{90.52}       & \multicolumn{1}{c|}{86.47}       & \multicolumn{1}{c|}{91.15}          & 86.30       & \multicolumn{1}{c|}{74.04}       & \multicolumn{1}{c|}{86.58}       & \multicolumn{1}{c|}{80.59}       & \multicolumn{1}{c|}{86.30}          & 82.98       \\ \hline \hline
\multirow{9}{*}{\textbf{RECALL}}    & 0.1                                 & \multicolumn{1}{c|}{36.08}       & \multicolumn{1}{c|}{39.60}          & \multicolumn{1}{c|}{35.91}       & \multicolumn{1}{c|}{39.63}       & 36.14       & \multicolumn{1}{c|}{40.19}       & \multicolumn{1}{c|}{40.51}       & \multicolumn{1}{c|}{40.08}       & \multicolumn{1}{c|}{40.59}          & 36.20       & \multicolumn{1}{c|}{39.49}       & \multicolumn{1}{c|}{40.39}       & \multicolumn{1}{c|}{39.40}       & \multicolumn{1}{c|}{40.48}          & 36.17       \\ \cline{2-17} 
                                    & 0.2                                 & \multicolumn{1}{c|}{39.20}       & \multicolumn{1}{c|}{47.11}          & \multicolumn{1}{c|}{38.98}       & \multicolumn{1}{c|}{47.76}       & 37.81       & \multicolumn{1}{c|}{51.19}       & \multicolumn{1}{c|}{52.16}       & \multicolumn{1}{c|}{51.19}       & \multicolumn{1}{c|}{52.50}          & 40.05       & \multicolumn{1}{c|}{49.52}       & \multicolumn{1}{c|}{52.16}       & \multicolumn{1}{c|}{49.32}       & \multicolumn{1}{c|}{51.90}          & 39.63       \\ \cline{2-17} 
                                    & 0.3                                 & \multicolumn{1}{c|}{42.55}       & \multicolumn{1}{c|}{54.57}          & \multicolumn{1}{c|}{41.95}       & \multicolumn{1}{c|}{54.54}       & 38.27       & \multicolumn{1}{c|}{65.33}       & \multicolumn{1}{c|}{68.42}       & \multicolumn{1}{c|}{65.42}       & \multicolumn{1}{c|}{69.47}          & 43.79       & \multicolumn{1}{c|}{61.28}       & \multicolumn{1}{c|}{69.08}       & \multicolumn{1}{c|}{61.39}       & \multicolumn{1}{c|}{68.25}          & 42.80       \\ \cline{2-17} 
                                    & 0.4                                 & \multicolumn{1}{c|}{43.96}       & \multicolumn{1}{c|}{62.81}          & \multicolumn{1}{c|}{41.38}       & \multicolumn{1}{c|}{64.94}       & 39.03       & \multicolumn{1}{c|}{78.46}       & \multicolumn{1}{c|}{85.29}       & \multicolumn{1}{c|}{78.63}       & \multicolumn{1}{c|}{86.99}          & 51.28       & \multicolumn{1}{c|}{69.25}       & \multicolumn{1}{c|}{83.31}       & \multicolumn{1}{c|}{71.00}       & \multicolumn{1}{c|}{83.96}          & 47.85       \\ \cline{2-17} 
                                    & 0.5                                 & \multicolumn{1}{c|}{50.65}       & \multicolumn{1}{c|}{60.46}          & \multicolumn{1}{c|}{46.80}       & \multicolumn{1}{c|}{70.78}       & 48.47       & \multicolumn{1}{c|}{82.97}       & \multicolumn{1}{c|}{93.28}       & \multicolumn{1}{c|}{84.75}       & \multicolumn{1}{c|}{\textbf{95.81}} & 61.76       & \multicolumn{1}{c|}{74.52}       & \multicolumn{1}{c|}{91.35}       & \multicolumn{1}{c|}{77.33}       & \multicolumn{1}{c|}{\textbf{93.08}} & 58.31       \\ \cline{2-17} 
                                    & 0.6                                 & \multicolumn{1}{c|}{58.96}       & \multicolumn{1}{c|}{68.31}          & \multicolumn{1}{c|}{56.01}       & \multicolumn{1}{c|}{78.01}       & 58.84       & \multicolumn{1}{c|}{83.36}       & \multicolumn{1}{c|}{94.30}       & \multicolumn{1}{c|}{86.93}       & \multicolumn{1}{c|}{95.10}          & 71.29       & \multicolumn{1}{c|}{75.94}       & \multicolumn{1}{c|}{91.55}       & \multicolumn{1}{c|}{81.97}       & \multicolumn{1}{c|}{92.23}          & 65.82       \\ \cline{2-17} 
                                    & 0.7                                 & \multicolumn{1}{c|}{62.76}       & \multicolumn{1}{c|}{81.24}          & \multicolumn{1}{c|}{68.91}       & \multicolumn{1}{c|}{81.46}       & 67.32       & \multicolumn{1}{c|}{81.69}       & \multicolumn{1}{c|}{93.34}       & \multicolumn{1}{c|}{89.37}       & \multicolumn{1}{c|}{93.34}          & 79.00       & \multicolumn{1}{c|}{78.71}       & \multicolumn{1}{c|}{90.73}       & \multicolumn{1}{c|}{86.03}       & \multicolumn{1}{c|}{90.62}          & 75.06       \\ \cline{2-17} 
                                    & 0.8                                 & \multicolumn{1}{c|}{60.40}       & \multicolumn{1}{c|}{82.48}          & \multicolumn{1}{c|}{71.74}       & \multicolumn{1}{c|}{80.24}       & 69.08       & \multicolumn{1}{c|}{84.47}       & \multicolumn{1}{c|}{92.63}       & \multicolumn{1}{c|}{90.87}       & \multicolumn{1}{c|}{92.60}          & 83.93       & \multicolumn{1}{c|}{81.09}       & \multicolumn{1}{c|}{90.19}       & \multicolumn{1}{c|}{86.42}       & \multicolumn{1}{c|}{88.66}          & 79.76       \\ \cline{2-17} 
                                    & 0.9                                 & \multicolumn{1}{c|}{61.90}       & \multicolumn{1}{c|}{\textbf{83.13}} & \multicolumn{1}{c|}{54.34}       & \multicolumn{1}{c|}{75.43}       & 64.54       & \multicolumn{1}{c|}{75.85}       & \multicolumn{1}{c|}{90.39}       & \multicolumn{1}{c|}{86.93}       & \multicolumn{1}{c|}{91.10}          & 85.06       & \multicolumn{1}{c|}{72.68}       & \multicolumn{1}{c|}{87.10}       & \multicolumn{1}{c|}{79.82}       & \multicolumn{1}{c|}{85.18}          & 81.69       \\ \hline \hline
\multirow{9}{*}{\textbf{RETRIEVAL}} & 0.1                                 & \multicolumn{1}{c|}{37.19}       & \multicolumn{1}{c|}{39.71}          & \multicolumn{1}{c|}{37.04}       & \multicolumn{1}{c|}{39.59}       & 36.59       & \multicolumn{1}{c|}{40.28}       & \multicolumn{1}{c|}{40.31}       & \multicolumn{1}{c|}{40.21}       & \multicolumn{1}{c|}{40.43}          & 36.73       & \multicolumn{1}{c|}{39.71}       & \multicolumn{1}{c|}{40.29}       & \multicolumn{1}{c|}{39.75}       & \multicolumn{1}{c|}{40.36}          & 36.68       \\ \cline{2-17} 
                                    & 0.2                                 & \multicolumn{1}{c|}{40.45}       & \multicolumn{1}{c|}{48.57}          & \multicolumn{1}{c|}{39.88}       & \multicolumn{1}{c|}{48.31}       & 38.68       & \multicolumn{1}{c|}{52.29}       & \multicolumn{1}{c|}{52.75}       & \multicolumn{1}{c|}{52.23}       & \multicolumn{1}{c|}{53.08}          & 39.62       & \multicolumn{1}{c|}{49.63}       & \multicolumn{1}{c|}{52.10}       & \multicolumn{1}{c|}{49.60}       & \multicolumn{1}{c|}{51.95}          & 39.04       \\ \cline{2-17} 
                                    & 0.3                                 & \multicolumn{1}{c|}{44.76}       & \multicolumn{1}{c|}{58.96}          & \multicolumn{1}{c|}{43.78}       & \multicolumn{1}{c|}{59.66}       & 39.03       & \multicolumn{1}{c|}{69.73}       & \multicolumn{1}{c|}{71.11}       & \multicolumn{1}{c|}{69.40}       & \multicolumn{1}{c|}{71.78}          & 44.85       & \multicolumn{1}{c|}{62.80}       & \multicolumn{1}{c|}{69.41}       & \multicolumn{1}{c|}{63.84}       & \multicolumn{1}{c|}{69.77}          & 43.84       \\ \cline{2-17} 
                                    & 0.4                                 & \multicolumn{1}{c|}{45.02}       & \multicolumn{1}{c|}{67.91}          & \multicolumn{1}{c|}{42.25}       & \multicolumn{1}{c|}{69.05}       & 39.78       & \multicolumn{1}{c|}{83.54}       & \multicolumn{1}{c|}{88.10}       & \multicolumn{1}{c|}{83.61}       & \multicolumn{1}{c|}{88.97}          & 53.47       & \multicolumn{1}{c|}{70.65}       & \multicolumn{1}{c|}{84.09}       & \multicolumn{1}{c|}{72.43}       & \multicolumn{1}{c|}{83.93}          & 49.02       \\ \cline{2-17} 
                                    & 0.5                                 & \multicolumn{1}{c|}{52.54}       & \multicolumn{1}{c|}{73.53}          & \multicolumn{1}{c|}{47.07}       & \multicolumn{1}{c|}{78.05}       & 47.30       & \multicolumn{1}{c|}{88.61}       & \multicolumn{1}{c|}{96.38}       & \multicolumn{1}{c|}{90.38}       & \multicolumn{1}{c|}{\textbf{97.53}} & 65.33       & \multicolumn{1}{c|}{74.45}       & \multicolumn{1}{c|}{91.74}       & \multicolumn{1}{c|}{77.89}       & \multicolumn{1}{c|}{\textbf{93.35}} & 56.88       \\ \cline{2-17} 
                                    & 0.6                                 & \multicolumn{1}{c|}{61.37}       & \multicolumn{1}{c|}{75.94}          & \multicolumn{1}{c|}{58.41}       & \multicolumn{1}{c|}{80.57}       & 60.15       & \multicolumn{1}{c|}{88.12}       & \multicolumn{1}{c|}{96.25}       & \multicolumn{1}{c|}{91.09}       & \multicolumn{1}{c|}{97.21}          & 76.46       & \multicolumn{1}{c|}{76.95}       & \multicolumn{1}{c|}{92.24}       & \multicolumn{1}{c|}{82.72}       & \multicolumn{1}{c|}{93.12}          & 65.81       \\ \cline{2-17} 
                                    & 0.7                                 & \multicolumn{1}{c|}{62.48}       & \multicolumn{1}{c|}{81.60}          & \multicolumn{1}{c|}{72.45}       & \multicolumn{1}{c|}{82.28}       & 69.65       & \multicolumn{1}{c|}{86.17}       & \multicolumn{1}{c|}{95.00}       & \multicolumn{1}{c|}{91.72}       & \multicolumn{1}{c|}{96.10}          & 82.96       & \multicolumn{1}{c|}{80.08}       & \multicolumn{1}{c|}{91.82}       & \multicolumn{1}{c|}{86.30}       & \multicolumn{1}{c|}{91.99}          & 75.04       \\ \cline{2-17} 
                                    & 0.8                                 & \multicolumn{1}{c|}{62.41}       & \multicolumn{1}{c|}{82.81} & \multicolumn{1}{c|}{74.39}       & \multicolumn{1}{c|}{80.87}       & 70.44       & \multicolumn{1}{c|}{88.24}       & \multicolumn{1}{c|}{94.18}       & \multicolumn{1}{c|}{93.03}       & \multicolumn{1}{c|}{94.49}          & 86.27       & \multicolumn{1}{c|}{83.00}       & \multicolumn{1}{c|}{90.60}       & \multicolumn{1}{c|}{87.46}       & \multicolumn{1}{c|}{89.49}          & 80.71       \\ \cline{2-17} 
                                    & 0.9                                 & \multicolumn{1}{c|}{63.44}       & \multicolumn{1}{c|}{\textbf{83.40}}          & \multicolumn{1}{c|}{55.68}       & \multicolumn{1}{c|}{77.96}       & 71.80       & \multicolumn{1}{c|}{78.56}       & \multicolumn{1}{c|}{92.25}       & \multicolumn{1}{c|}{88.52}       & \multicolumn{1}{c|}{92.79}          & 87.62       & \multicolumn{1}{c|}{74.87}       & \multicolumn{1}{c|}{87.43}       & \multicolumn{1}{c|}{79.86}       & \multicolumn{1}{c|}{86.92}          & 83.03       \\ \hline
\end{tabular}}
\label{T1:CPTE}
\vspace{0.15cm}
\scriptsize{\\
Note: the bold entries represent the highest classification accuracy percentage obtained by each classifier for different events.\\
f1 - clustering coefficient, f2 - eigenvector centrality, f3 - betweenness, f4 - degrees, f5 - coreness centrality}
\end{table*}

\begin{figure*}[t]
  \centering
  \includegraphics[width = 0.95\linewidth]{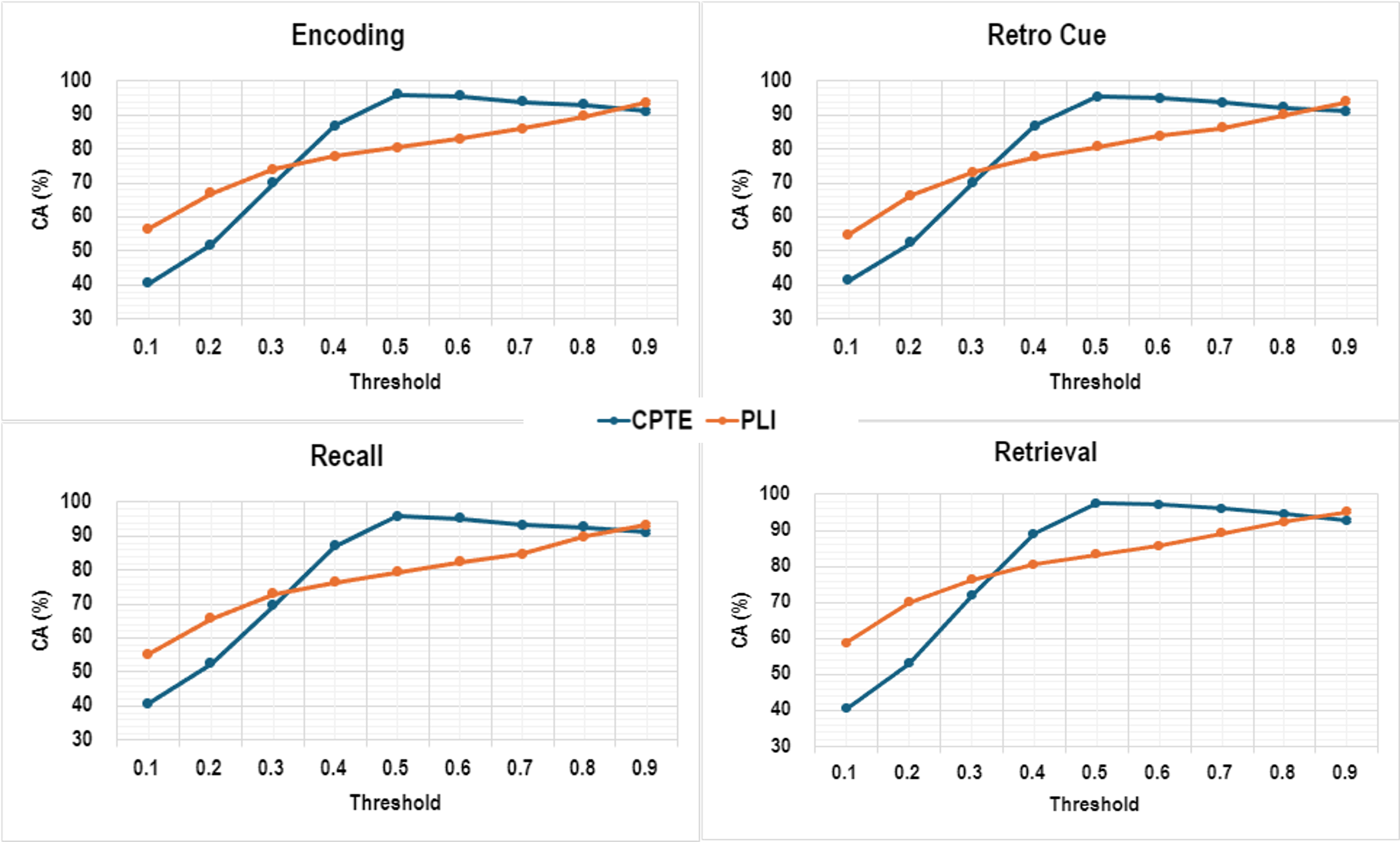}
  \caption{The plots illustrate the complex network organization analysis computed using different methods utilizing the Degree feature and the Random Forest (RF) classifier. The CPTE-based complex network analysis yielded the highest classification accuracy in distinguishing between AD, MCI, and HC groups.}
  \label{fig:CPTE_Plots}
\end{figure*}
\subsection{Network Parameters Extraction}\label{subsec:network_para}

This section elaborates on the network parameters derived from the Network Organization Matrices (NOMs) to analyze brain network dynamics in study groups. These parameters provide quantitative measures of functional connectivity and were subsequently used as input features for machine learning-based dementia classification. Five key network metrics were computed for each WM stage: \(D\), \(CC\), \(EC\), \(BC\), and \((C^c)\) for the sensor as a node, as detailed below.

\subsubsection{Degree (\(D\))}
The degree (\(D\)) measures the number of connections (or edges) a node has within the network. It quantifies the local connectivity of a node, representing its potential influence within the network. A high degree indicates that the node facilitates rapid information dissemination, while a low degree suggests a peripheral role with limited influence within the network structure. Let $A$ denote the adjacency matrix (or NOM) of size $I \times I$, where each entry 
$A_{v_kv_j}$ specifies the connection strength between nodes $v_k$ and $v_j$. The degree of a node $k$ is formally defined as:  

\begin{equation}
D_k = \sum_{j=1}^I A_{v_kv_j}, \quad k = 1, \dots, I \tag{9}
\end{equation}
 The degree $D_k$ thus reflects either the total number of connections or the cumulative strength of connections for node $k$.

\subsubsection{Clustering coefficient (\(C\))}

The clustering coefficient (\(C\)) is a metric that quantifies the extent to which a node neighborhood forms a cohesive group (cluster) in the graph. It provides insight into the local interconnectivity of the network. For a given node \( k \), the clustering coefficient \( C_{k} \) is defined as:

\[
C_{k} = \frac{2E_{k}}{D_{k}(D_{k} - 1)} \quad k = 1, \dots, I \tag{10}
\]

where, \( E_{k} \) is the number of edges connecting the neighbors of node \( k \), \( D_{k} \) is the degree of node \( k \). The value of C ranges from 0-1, where a value of 1 indicates that all neighbors of the node are interconnected while a value of 0 indicates that none of the neighbors are connected.

\subsubsection{Eigenvector Centrality (\(EC\))}

Eigenvector centrality (EC) measures the importance of a node within the network by considering the connection of the node to other densely connected nodes. A higher EC indicates a crucial node in the network. The eigenvector centrality \( EC_{k} \) of node \( k \) is defined as:

\[
EC_{k} = \frac{1}{\lambda} \sum_{v_j=1}^{I} A_{v_k v_j} EC_{v_j}\tag{11}
\]

Here, \( \lambda \) is a constant, and \( A \) represents the connectivity matrix. In matrix form:

\[
\lambda x = A \cdot x\tag{12}
\]

Thus, \( x \) is an eigenvector of \( A \) with eigenvalue \( \lambda \). According to the Perron-Frobenius theorem, \( \lambda \) must be the largest eigenvalue for non-negative centralities.


\subsubsection{Betweenness Centrality (BC)}

Betweenness centrality (\(BC\)) quantifies the extent to which a node lies on the shortest paths between other nodes. It measures the influence of nodes within a network by reflecting their role as a bridge between other nodes. The mathematical formulation of \(BC\) of a node $k$ is defined as:  

\begin{equation}
BC_k = \frac{1}{(I - 1)(I - 2)} 
\sum_{\substack{v_h,v_j \in V \\ h \neq j, \, h \neq k, \, j \neq k}} 
\frac{\rho_{v_hv_j}(v_k)}{\rho_{v_hv_j}},
\end{equation}

where $I$ is the total number of nodes in the network $V$ (corresponding to the number of electrodes $I$), $\rho_{v_hv_j}$ denotes 
the total number of shortest paths between nodes $h$ and $j$, and $\rho_{hj}(k)$ is the number of those paths that pass through node $k$. The normalization term $(I-1)(I-2)$ ensures comparability across networks of different sizes. Nodes with larger $BC_k$ values are considered crucial for maintaining efficient global integration, as they mediate communication between otherwise distant or weakly connected subnetworks.








\subsubsection{Coreness Centrality \((C^c)\)}
 
This measure \cite{yi2016identifying, bae2014identifying} quantifies the hierarchical position of a node within the core–periphery structure of the network. Nodes with high $C^c$ values belong to the central core, reflecting their critical role in global information integration. In contrast, nodes with low $C^c$ values are positioned toward the periphery, where they primarily support localized information processing. Coreness centrality is computed using $s$-shell decomposition. Formally, the coreness centrality of a node $k$ is defined as:  

\begin{equation}
C^c_k = \sum_{j \in \Gamma(v_k)} s(v_j),
\end{equation}

where $\Gamma(v_k)$ denotes the set of neighbors directly connected to node $v_k$, and $s(v_j)$ is the shell index assigned to node $j$. The shell index $s(v_j)$ is determined by iteratively removing nodes with a degree less than or equal to a given threshold, thereby identifying nested layers of the network. Nodes with larger $C^c_k$ values are thus more deeply embedded in the structural core of the network, while nodes with smaller values occupy peripheral layers with limited global influence.


The network parameters were computed for NOMs derived from epochs of \([V]\), \([V^{SH}_{nm}]\), and \([V^{H^2}_{nm}]\) for each WM stage and study group. A feature vector of size 63 was obtained for each computed NOM with individual network parameters (\(D\), \(C\), \(EC\), \(BC\), or \(C^c\)). These features were utilized as inputs for supervised machine learning models, specifically random forest (RF), support vector machine (SVM), and extreme gradient boosting (XGBoost), for dementia classification across WM stages. For performance evaluation, classification accuracy was assessed using a 10-fold cross-validation technique.


\begin{figure*}[ht]
    \centering
    \adjustbox{trim= 0 0cm 0 0}{%
    \includegraphics[width = 0.95\linewidth]{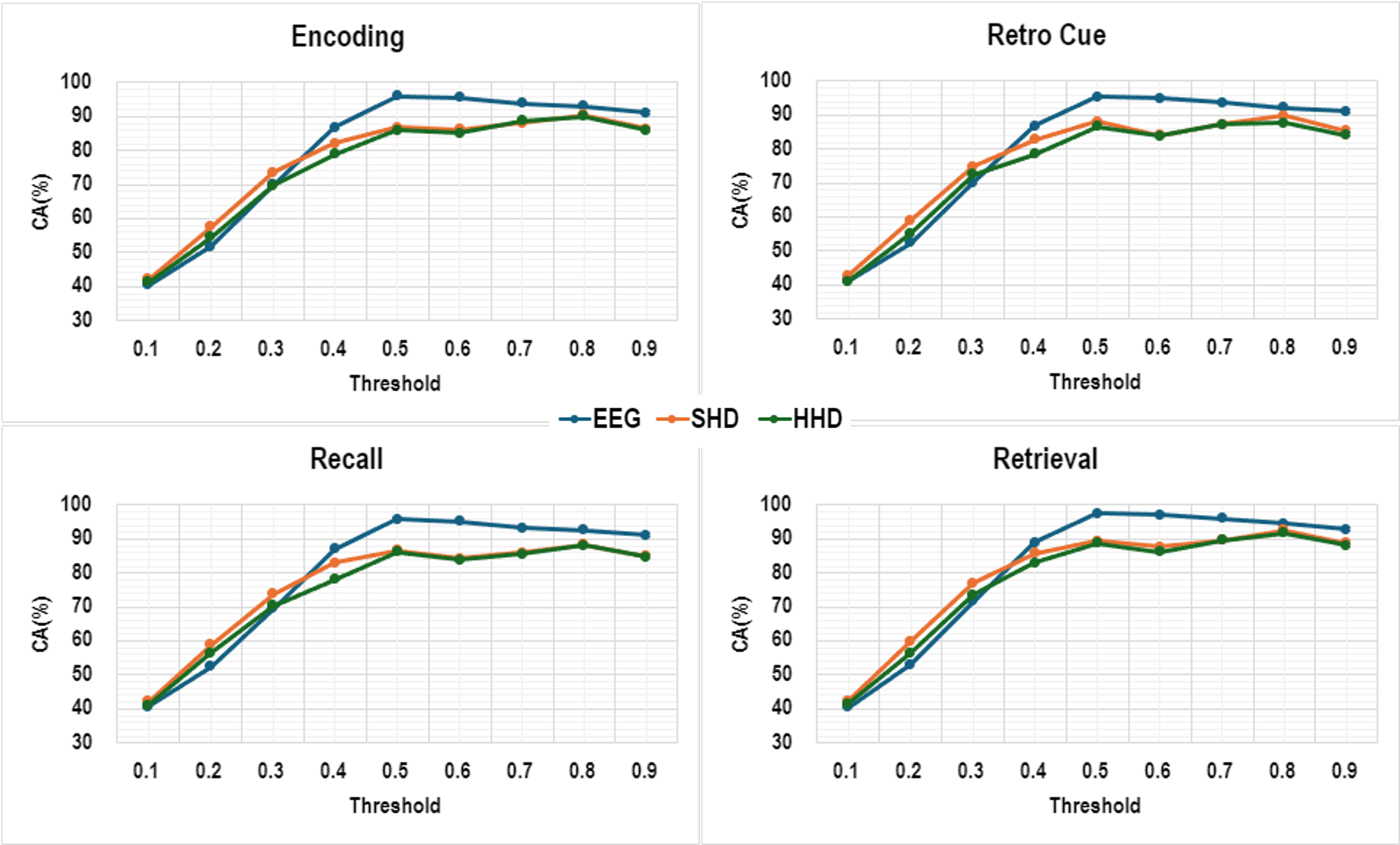}}
    \caption{The plots show the variation in classification accuracy of the RF classifier with Degree network parameter for different features of EEG signals. It illustrates CPTE-based network organization across thresholds for encoding, retro cue, recall, and retrieval stages of vWM. The classification is performed for AD, MCI, and HC groups. }
    \label{fig:EEG_HH_SH}
\end{figure*}

\section{Results}\label{sec:results}

This section presents the performance evaluation of the proposed framework with respect to various parameters to classify dementia stages across working memory (WM) tasks. The processed EEG signals \([V]\), spherical harmonics \([V^{SH}_{nm}]\), and head harmonics \([V^{H^2}_{nm}]\) were utilized to extract 1 s epochs of each WM stage after their onset for all participants. Each epoch was further segmented into 0.5 s windows with 50\% overlap, resulting in segments of 2400, 1200, 1200, and 4800 for encoding, retro-cue, recall, and retrieval, respectively in each study group. For each segment, NOMs were computed using PLI and CPTE methods, followed by standardization of the NOMs elements in the range of $[0,1]$ and binarization using varying threshold values, as detailed in Section \ref{subsec:NOM_compute}. Further, the binarized NOMs were utilized for network parameters (D, C, EC, BC, and \(C^c\)) extraction. The classification performance was evaluated using machine learning models, including RF, SVM, and XGBoost, with classification accuracy percentage (CA(\%)) as a performance metric. The comparative classification analysis was performed for different spatiotemporal EEG features, NOMs, binarization threshold values, and network parameters for various WM stages.  

\begin{figure*}[ht]
    \centering
    \includegraphics[width = \linewidth]{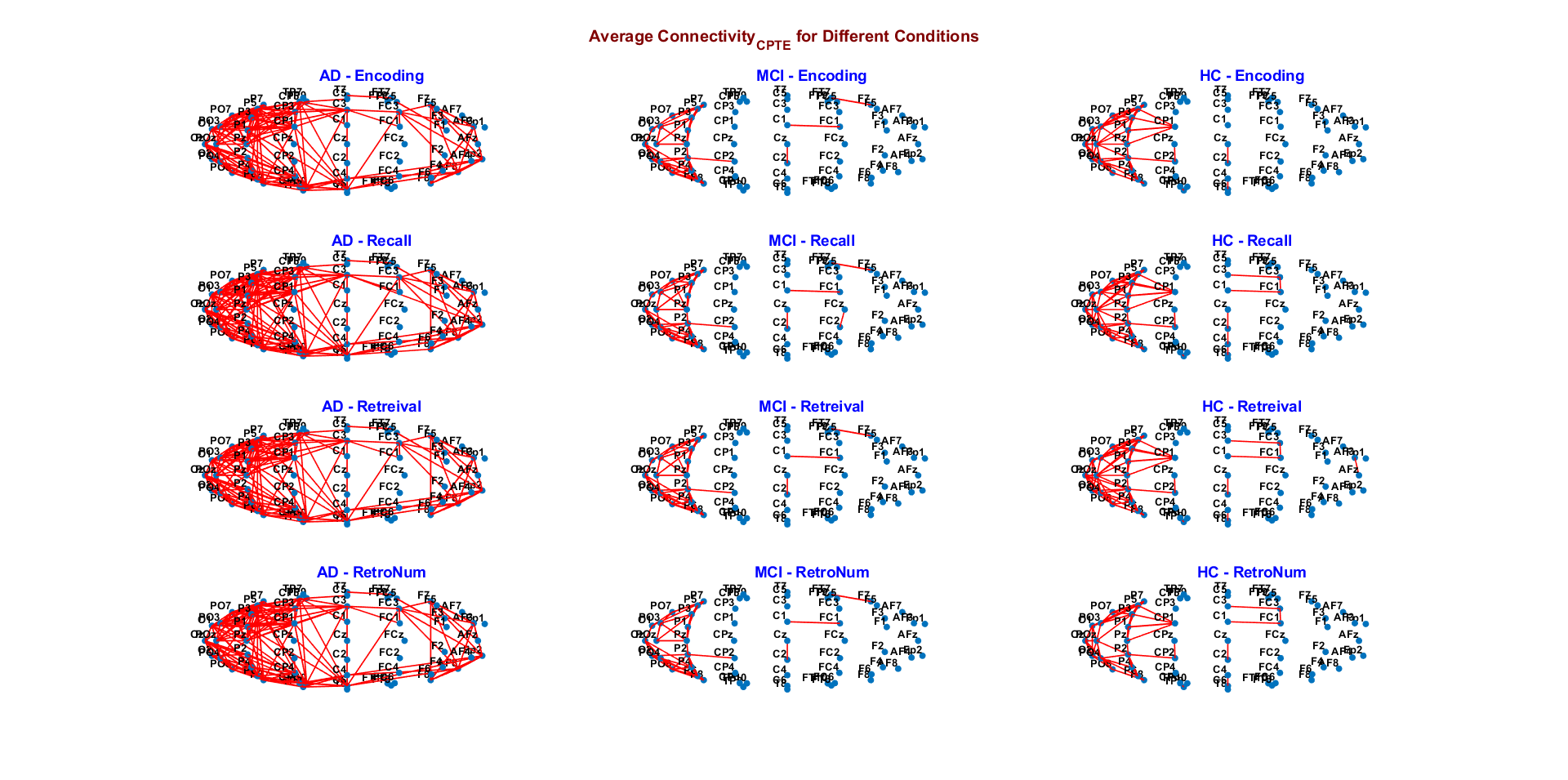}
    \caption{Connectivity plots based on CPTE for all WM stages and groups with a 0.4 threshold value.}
    \label{fig:Connectivity_CPTE}
\end{figure*}

Table \ref{T1:CPTE} shows the classification performance using CPTE-based network parameters for different threshold values and classifiers for temporal EEG features \([V]\). In particular, the highest classification accuracy of 97.53\% is obtained for the retrieval WM stage with a threshold value of 0.5 using the degree (\(D\)) network parameter and RF classifier. Figure \ref{fig:CPTE_Plots} shows a performance comparison of CPTE and PLI-based NOMs using \(D\) network parameters with RF classifier for different WM stages. It can be noted that the CPTE-based network feature outperformed the PLI-based approach in distinguishing AD, MCI, and HC groups across WM stages. Further, a comparison analysis using CPTE-based NOM with \([V]\), \([V^{SH}_{nm}]\), and \([V^{H^2}_{nm}]\) EEG features is shown in Figure \ref{fig:EEG_HH_SH} for all WM stages. In particular, network parameter \( D \) is utilized with an RF classifier for the analysis. Also, decomposition order \( N =6\) is used for extracting spatio-temporal harmonics EEG features. The connectivity plots for all groups and WM stages are presented in Figure \ref{fig:Connectivity_CPTE}. It may be noted that the connectivity of the AD group is higher than that of the MCI and HC groups across every WM stage in the CPTE-based complex network. 



\section{Discussion}\label{sec:discuss}


\subsection{Comparison between CPTE and PLI NOMs}
In this study, CPTE and PLI-based NOMs were explored for dementia stage classification for WM tasks. It can be noted that CPTE-based network parameters outperformed the PLI-based method to distinguish between AD, MCI, and HC groups across all WM tasks, as shown in Figure \ref{fig:CPTE_Plots}. CPTE-based RF classifier with degree network parameters demonstrated consistently higher accuracy beyond a threshold value of 0.5 in comparison to the PLI-based approach across the WM stages. The RF classifier achieved the highest CA for CPTE-based degrees network parameters for a threshold value of 0.5, as shown in Table \ref{T1:CPTE}. In particular, the highest CA of 96.10\%, 95.43\%, 95.81\%, and 97.53\% were obtained using degree network parameters with the RF classifier for encoding, retro-cue, recall, and retrieval, respectively. The corresponding PLI-based analyses are presented in Supplementary Table I.


Inter-channel connectivity plots averaged across all epochs for each group are presented in Figure \ref{fig:Connectivity_CPTE} to visualize the classification heuristics. It can be observed that connectivity in the AD group is significantly higher than in the MCI and HC groups across the WM stages. This trend aligns with the characteristic hypoactivity of neurodegenerative conditions like AD during WM stages, reflecting reduced complexity \cite{sun2020complexity} and increased channel synchronization. In contrast, the HC group exhibits selective synchronization among specific channels, potentially relevant to WM tasks. This connectivity could reflect an efficient network that supports task-relevant processing. The results indicate that patients with neurodegenerative conditions exhibited impairment during the WM task \cite{lou2015decreased} and showed higher synchronized network organization compared to healthy controls. PLI-based connectivity analysis is shown in Supplementary Figure 7. 

\subsection{Network Parameters Analysis}
This study included five network parameters \( D \), \( C \), \( EC \), \( BC \), and \( C^c \) for dementia classification, as shown in Table\ref{T1:CPTE}. Network parameters \( D \) and  \( EC \) had better classification performance compared to \( C \), \( BC \), and \( C^c \). This can be attributed to the fact that degree (\( D \)) and eigenvector centrality (\( EC \)) capture both local and global connectivity patterns. In particular, \( D \) was effective in distinguishing network alterations in AD, MCI, and HC study groups. The peak classification accuracies were obtained using the RF classifier with \( D \) network parameter, followed by XGBoost, as shown in Table \ref{T1:CPTE}. However, the SVM classifier yielded the best classification performance using the \( EC \) network parameter.


\subsection{Significance of Retrieval stage in Dementia classification}
The retrieval stage among WM tasks emerged as the most informative to classify between AD, MCI, and HC groups, achieving the highest CA of 97.53\% among all WM tasks. The retrieval stage is critical to understand the information processing and the causality relationships among brain regions \cite{jiang2024exploration}. The encoding, retro-cue, and recall stages also yielded high classification performance, emphasizing their roles in characterizing WM processes.

\subsection{Comparative analysis of different  spatiotemporal EEG features}
To account for the spatial information of the head geometry \cite{perrin1989spherical}, the SHD and HHD techniques were explored as inputs for complex network analysis and dementia classification during the WM stages. As shown in Figure \ref{fig:EEG_HH_SH}, the classification performance achieved using EEG \([V]\) spatiotemporal feature outperformed SHD (\([V^{SH}_{nm}]\)), and HHD (\([V^{H^2}_{nm}]\)). While SHD and HHD offer computational efficiency and dimensionality reduction, their performance was affected by information loss inherent in feature compression. The \([V]\) preserved the full dimensionality of EEG signals, resulting in higher classification accuracy. However, SHD and HHD demonstrated reasonable performance, with the highest accuracy of 92.53\% and 91.78\% in retrieval, 90.54\% and 89.97\% in encoding, 88.35\% and 88.04\% in recall, and 89.84\% and 87.8\% in retro cue, respectively. A detailed analysis for varying order $N$ across different thresholds and classifiers, is presented in Supplementary Table II.

\subsection{Effect of Thresholding}
Applying thresholding to binarize the Network Organization Matrices (NOMs) significantly improved the interpretability of connectivity metrics by isolating stronger connections and discarding the weaker ones. This process effectively highlighted the most critical features of the brain's network organization, enabling a more precise analysis of connectivity patterns. As shown in Table \ref{T1:CPTE}, the threshold of 0.5 consistently produced the highest classification accuracy across the WM stages. Similar observations can be depicted in Figure \ref{fig:CPTE_Plots} and Figure \ref{fig:EEG_HH_SH}. These findings emphasize the critical role of threshold optimization in refining connectivity metrics for ensuring robust performance and interpretation of neurophysiological data.

\subsection{Statistical Analysis}
 Two-way ANOVA for AD, MC, and HC revealed a significant main effect of group ($p < 0.01$), WM stage ($p < 0.05$), as well as a significant group-stage interaction ($p < 0.01$) on CPTE-based network features. These findings highlight both disease-specific and stage-dependent alterations in functional connectivity. 
\section{Conclusion}\label{sec:conclusion}
This study highlights the pivotal role of working memory (WM) tasks in distinguishing the different stages of neurodegenerative diseases such as AD and MCI. A peak classification accuracy of 97.53\% was achieved using EEG-based CPTE connectivity measure, particularly in the retrieval phase. This underscores WM's diagnostic potential in early detection of cognitive decline. CPTE approach emerged as a robust method for capturing brain network dynamics and assessing disruptions caused by neurodegeneration, demonstrating significant improvements compared to traditional metrics such as the PLI. The findings of the study indicate that neurodegenerative diseases show increased synchronized network organization in WM processes compared to healthy controls, reflecting compensatory mechanisms for cognitive deficits. Network parameters like \(D\) and \(EC\) emerged as the most effective features, emphasizing their relevance in detecting disrupted network organization in dementia. While spatiotemporal features derived from spherical and head harmonics decomposition offer computational efficiency, full-dimensional spatiotemporal features delivered higher classification accuracy. Thresholding improved the interpretability of connectivity metrics, with a 0.5 threshold yielding the best classification performance. The findings of this study establish CPTE-based EEG analysis as a powerful tool for understanding and classifying dementia stages. It offers a scalable and cost-effective solution for early diagnosis and monitoring of neurodegenerative diseases. These advancements have the potential to significantly improve clinical outcomes and the quality of life for individuals affected by dementia. Future research could validate the methodology across larger datasets and more diverse neurodegenerative populations.

\section*{Declarations}

\subsection*{Consent for publication}
 Not applicable
 
\subsection*{Ethics approval and consent to participate}
This work involved human subjects in its research. Approval
of all ethical and experimental procedures and protocols was granted by the
Institute Ethics Committee, All India Institute of Ayurveda, New Delhi, India, with Ref No. IEC-331/27.06.2023/Rp(E)-12/2023, dated: 17/08/2023. All participants provided informed consent prior to the study.

\subsection*{Availability of data and material}
The datasets utilized or analyzed in the present study will be accessible upon reasonable request from the corresponding
author.

\subsection*{Competing interests}
The authors declare that they have no competing interests.

\subsection*{Funding}
This work was supported in part by IIT Mandi iHub and HCI Foundation
India with project number RP04502G.

\subsection*{Acknowledgment}
The authors would like to thank Prof. Pradeep Kumar Prajapati, Vice Chancellor from Dr. Sarvepalli Radhakrishnan Rajasthan Ayurved University (DSSRAU), Jodhpur, and Dr. Lokesh Shekhawat, Assistant Professor from Department of Psychiatry, Atal Bihari Vajpayee Institute of Medical Sciences (ABVIMS), and Dr. Ram Manohar Lohia Hospital, New Delhi, for their expert discussion and guidance during the diagnosis and intervention processes.
Lastly, the authors extend special thanks to Dr. G.P. Bhagat, founder of SHEOWS Guru Vishram Vridh Ashram, for his support in facilitating the availability of the patients.

\subsection*{Authors’ contributions}
Conceptualization by S.R., A.J., P.Y., and L.K.; Methodology, Software, Formal analysis, Writing – original draft by S.R.; Data curation by S.R., H.S., and A.J.; Data acquisition by S.R., H.S., R.B., and A.K.; Clinical diagnosis by R.B. and P.Y.; Writing – review $\&$ editing by A.J., R.B., L.K., P.K., D.J. and Supervision by L.K., P.K. and D.J. All authors critically reviewed the manuscript.

\subsection*{Authors' information}
$^1$Department of Electrical Engineering, Indian Institute of Technology Delhi, India.
$^2$Department of RS and BK, All India Institute of Ayurveda Delhi, India.
$^3$Centre for Biomedical Engineering, Indian Institute of Technology Delhi, India.
$^4$Bharti School of Telecommunication and
Yardi School of Artificial Intelligence, Indian Institute of Technology Delhi, India.

\balance
\bibliographystyle{ieeetr}  
\bibliography{main_v1}


 \end{document}